\documentclass{kluwer}    

\usepackage{epsfig}
\usepackage{graphicx}

\newdisplay{guess}{Conjecture}

\begin{document}                                                                                   
\begin{article}
\begin{opening}         
\title{MHD simulations of the long-term evolution of a dipolar 
       magnetosphere surrounded by an accretion disk} 
\author{Christian \surname{Fendt}}  
\runningauthor{Ch.~Fendt}
\runningtitle{Dipolar magnetosphere --- MHD simulations}
\institute{Institut f\"ur Physik, Universit\"at Potsdam,
           Am Neuen Palais 10, D-14469 Potsdam,
           Germany; \quad E-mail: cfendt@aip.de}
\date{October 25, 2002}

\begin{abstract}
The evolution of an initially stellar dipole type magnetosphere
interacting with an accretion disk is investigated using
numerical ideal MHD simulations.
The simulations follow several 1000 Keplerian periods of
the inner disk (for animated movies see 
{\tt http://www.aip.de/$\sim$cfendt}).
Our model prescribes a Keplerian disk around a rotating
star as a fixed boundary condition. 
The initial magnetic field distribution remains frozen 
into the star and the disk. 
The mass flow rate into the corona is fixed for both components.

The initial dipole type magnetic field develops into a spherically
radial outflow pattern with two main components --- a disk wind and a 
stellar wind --- both evolving into a quasi-stationary final
state.
A neutral field line divides both components, along which
small plasmoids are ejected in irregular time intervals.
The half opening angle of the stellar wind cone varies from $30^o$ to
$55^o$ depending on the ratio of the mass flow rates of disk
wind and stellar wind.
The maximum speed of the outflow is about the Keplerian speed at the
inner disk radius.

An axial jet forms during the first decades of rotations.
However, this feature does not survive on the very long time scale
and a pressure driven low velocity flow along the axis evolves.
Within a cone of $15^o$ along the axis the formation of
knots may be observed if the stellar wind is weak.
With the chosen mass flow rates and field strength we see
almost no indication for a flow self-collimation.
This is due to the weak net poloidal electric current in the
magnetosphere which is in difference to typical jet models.
\end{abstract}

\end{opening}           

\def\ri{r_{\rm i}}
\def\ro{r_{\rm out}}
\def\rs{r_{\star}}
\def\mj{\dot{M}_{\rm jet}}
\def\ms{\dot{M}_{\star}}
\def\vk{v_{\rm K}}
\def\ti{t_{\rm i}}
\def\rj{R_{\rm jet}}
\def\rk{R_{\kappa}}
\def\bp{B_{\rm P}}
\def\bh{B_{\phi}}
\def\jp{\vec{B}_{\rm P}}
\def\cm{{\rm cm}}
\def\msun{{\rm M}_{\sun}}
\def\rsun{{\rm R}_{\sun}}

\section{Introduction}

A stellar dipole magnetic field surrounded by an accretion
disk is the model scenario for a variety of astrophysical sources
 --- classical T\,Tauri stars, cataclysmic variables or
high mass X-ray binaries.
Some of them exhibit Doppler shifted emission lines indicating wind
motion.
Highly collimated jets are observed from young stellar objects
and X-ray binaries.
In general, magnetic fields are thought to play the leading role
for jet acceleration and collimation
(Blandford \& Payne 1982; Camenzind 1990;
Shu et al. 1994; Fendt et al. 1995).

Several papers numerically consider the evolution of a stellar 
magnetic dipole interacting with a diffusive accretion disk
(Hayashi et al. 1996;
Miller \& Stone 1997; 
Goodson et al. 1997, 1999).
In these papers a collapse of the inner disk is indicated.
The inward accretion flow develops a shock near the star.
The stream becomes deflected resulting in a high-speed flow in axial
direction.
The results of Goodson et al. are especially interesting as combining
a huge spatial scale with high spatial resolution near the star.
However, to our understanding, it is not clear, how the initial condition
(a standard $\alpha$-viscosity disk) is actually evolving in their code
without physical viscosity.

We emphasize that time-dependent simulations lasting only a short
time period strongly depend on the initial condition.
The simulation of jet magnetosphere over
{\em many} rotational periods is essential.
The observed kinematic time scale of protostellar jets
can be as large as $10^4$\,yrs (or $5\times 10^5$ stellar 
rotations).
Thus, we follow another approach for the simulation of magnetized
winds from accretion disks and consider 
the accretion disk ``only'' as a {\em boundary condition}.
Since the disk structure itself is not treated, such simulations may last
over hundreds of Keplerian periods.
For a certain initial magnetic field distribution, a stationary state
self-collimating jet flow can be obtained 
(Ouyed \& Pudritz 1997, Fendt \& Cemeljic 2002).

In our project (Fendt \& Elstner 1999, 2000), we are essentially interested
in the evolution of the ideal magnetohydrodynamic (MHD) magnetosphere
and the formation of winds and jets and not in the evolution of the disk
structure itself.
Therefore, we do not include magnetic diffusivity into our simulations.
We follow the approach of Ouyed \& Pudritz (1997) in combination with
dipole magnetic field initial condition.
Using the ZEUS-3D code (Stone \& Norman 1992a,b) in the axisymmetry option
we solve the time-dependent ideal MHD equations,
$$
\frac{\partial \rho}{\partial t} + \nabla \cdot (\rho \vec{v} ) = 0\,, 
\quad\quad
\frac{\partial\vec{B} }{\partial t} -
\nabla \times (\vec{v} \times \vec{B}) = 0\,,
\quad\quad
\nabla \cdot\vec{B} = 0\,,
$$ 
$$ 
\rho \left[ \frac{\partial\vec{v} }{\partial t}
+ \left(\vec{v} \cdot \nabla\right)\vec{v} \right]
+ \nabla (P + P_{\rm A}) + \rho\nabla\Phi - \vec{j} \times \vec{B} = 0\,,
$$ 
where $\vec{B}$ is the magnetic field,
$\vec{v}$ the velocity,
$\rho $ the gas density,
$P $ the gas pressure,
$\vec{j} = \nabla \times \vec{B} / 4\pi$ the electric current density.
and $\Phi$ the gravitational potential.
We apply a polytropic equation of state (polytropic index $\gamma=5/3$)
and do not solve the energy equation.
Instead, the internal energy is defined given $ e=p/(\gamma-1)$.
Similar to Ouyed \& Pudritz (1997), 
we introduce an Alfv\'en wave turbulent magnetic pressure,
$P_{\rm A}\equiv P/\beta_T $,
with constant $\beta_T $.

The main parameters of our simulation are the
plasma beta just above the inner disk radius $r_i$,
$\beta_i \equiv 8 \pi P_i / B_i^2 $,
and the Mach number of the rotating gas,
$\delta_i \equiv \rho_i v^2_{K,i} / P_i $,
where $v_{K,i} \equiv \sqrt{GM/r_i} $.

\section{The model -- numerical realization}

Our model setup represents a central star surrounded by a Keplerian 
disk with a gap in between.

The initial coronal density distribution is in hydrostatic equilibrium.
The gravitational potential is not smoothed.
We choose the initial field distribution of a force-free, current-free
stellar magnetic dipole,
deformed by the effect of ``dragging'' in the disk.
Therefore the poloidal field is inclined towards the disk surface. 
Since the boundary condition for the poloidal magnetic field along
the inflow boundary is fixed,
the magnetic flux from the star and disk is conserved.
The disk toroidal magnetic field is also force-free,
$B_{\phi} = {\mu}_i / r $ for $r\geq r_{\rm i}$ with 
$\mu_i=B_{\phi,{\rm i}}/B_{\rm i}$.

Hydrodynamic boundary conditions are `inflow' along the $r$-axis,
`reflecting' along the symmetry axis
and `outflow' along the outer boundaries.
The inflow parameters into the corona are defined with respect to
the three different boundary regions -- star, gap and disk.
The stellar wind boundary condition is motivated by the fact
that {\em stellar winds} are indeed observed.
The ratio of mass flow rates in the two outflow components
also governs the structure of the flow.

\section{Results and discussion}

In the following we discuss the results of our simulations.
For details see Fendt \& Elstner (1999, 2000).
As a general behavior, the initial dipole type structure of the
magnetic field disappears on spatial scales larger than the
inner disk radius and a two component wind structure -- a disk
wind and a stellar wind -- evolves.
On the very long time scale we find a {\em quasi-stationary} final
state of a spherically radial mass outflow.

The stellar rotational period is chosen as
$\Omega_{\star} = (v_{K,i}/r_i) = 1$.
Thus, the magnetospheric co-rotation radius is at $r_i$.
Other parameters are 
$\beta_i = 1.0 $, $\delta_i= 100$, $\mu=-1.0$, and a stellar
radius $\rs = 0.5$.
A numerical grid of $250^2$ elements is used for
a box of physical size $20\times20 r_i$.
The stellar wind mass flow rate is rather large, 
$\dot{M}_{\star}/ \dot{M}_{\rm D} = 2$. 

\begin{figure} 
\includegraphics[width=12cm]{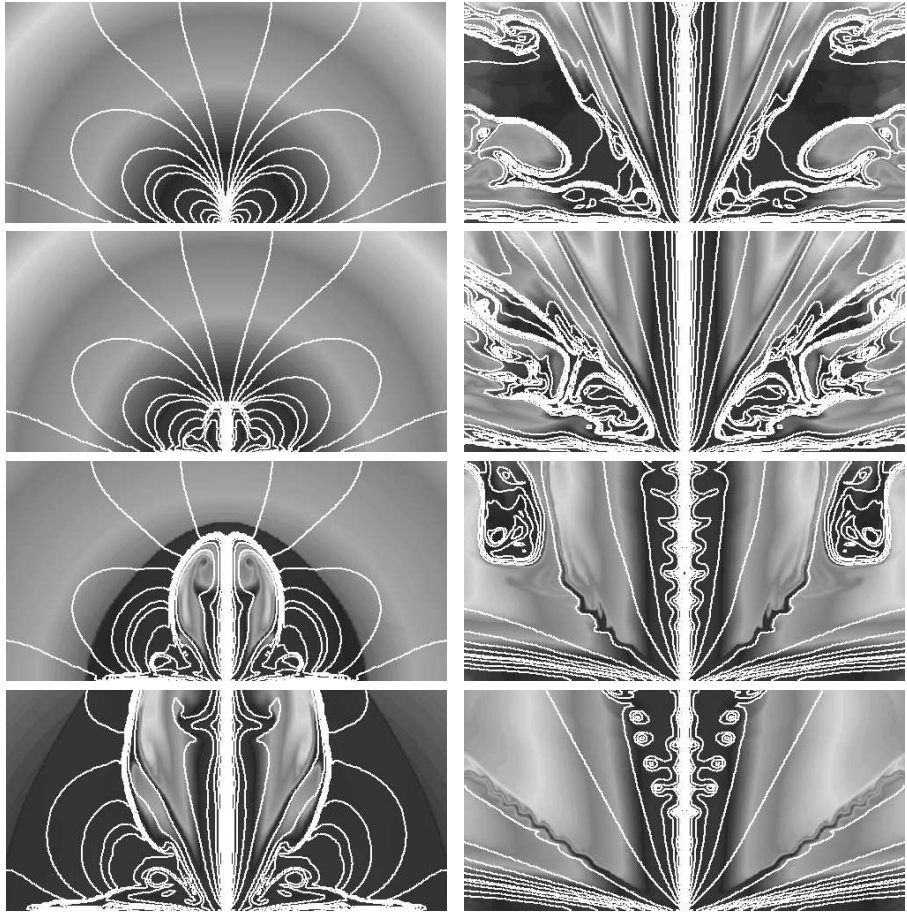}
\caption[]{Long-term evolution of a axisymmetric dipolar magnetosphere. 
Shown is density (grey scale) and poloidal magnetic field (white lines) 
for 0, 10, 25, 50, 100, 250, 500, 950, and 2700 rotations.
The symmetry axis points upwards.
}
\label{fig1}
\end{figure}

\begin{figure} 
\includegraphics[width=11cm]{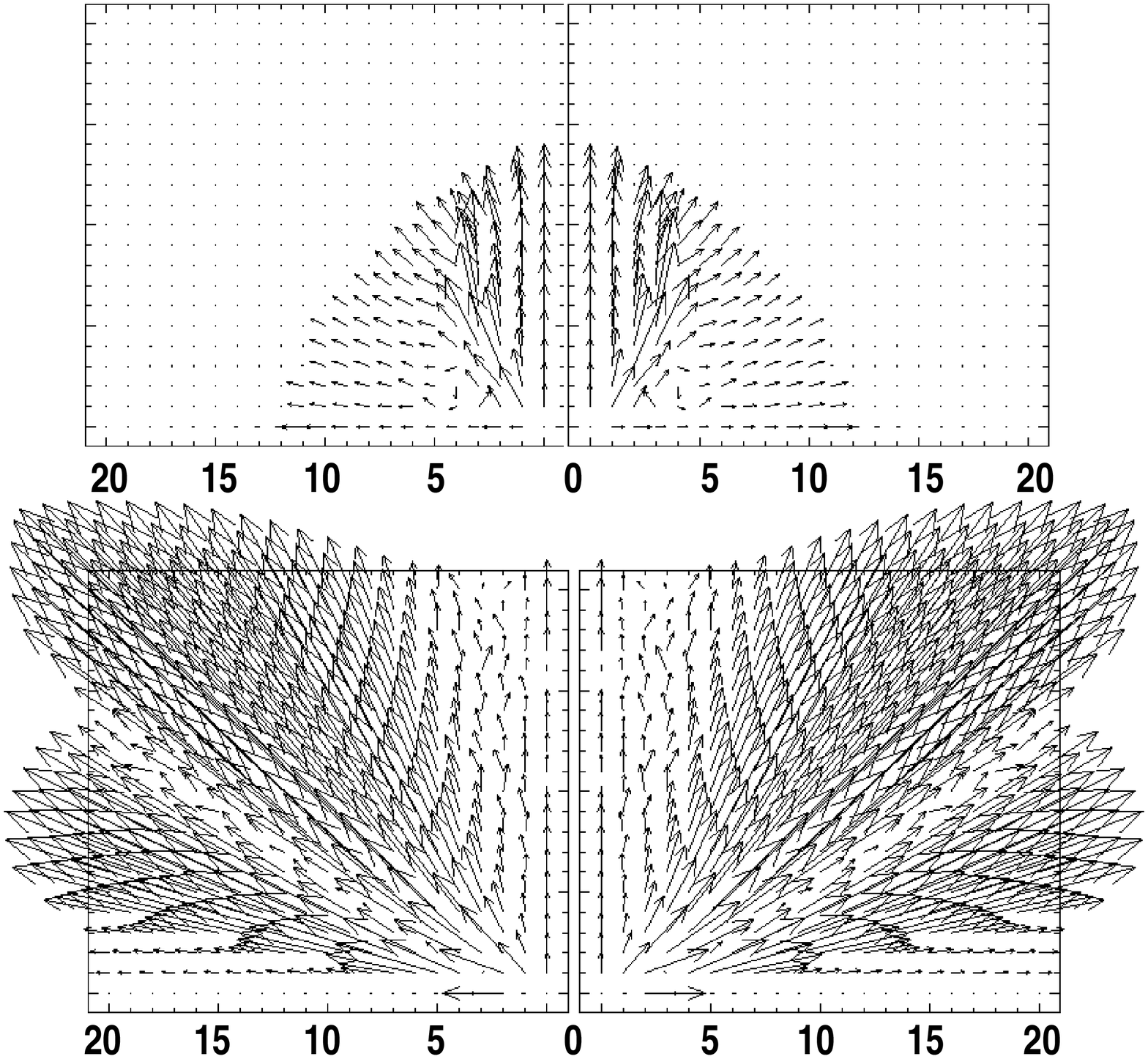}
\caption[]{Long-term evolution of a dipolar magnetosphere. 
The velocity vectors for the time step of 30 and 2500 rotations clearly
show the two flow components and the weakening of the axial jet.
The components are separated by the neutral field line.}
\label{fig1}
\end{figure}

\subsection{The first evolutionary stages}
During the first evolutionary stages the magnetospheric structure is
characterized by the following features.
The {\em winding-up} of the dipolar poloidal field,
the formation of a {\em neutral field line},
a {\em transient axial jet} feature,
a two component outflow consisting of
a {\em stellar wind} and a {\em disk wind}.
 
The winding-up process of poloidal magnetic field due to differential
rotation between star and disk and the static initial corona
induces a toroidal field with a positive (negative) sign along the
field lines outside (inside) the slowly emerging neutral field line.

During the first 50 rotations a jet feature evolves along the rotational
axis with a pattern velocity of about $0.2\,v_{\rm K,i}$.
Such an axial jet is known as a characteristic result of MHD simulations
performed in the recent literature 
(Hayashi et al. 1997; Goodson et al. 1997, 1999; Kudoh et al. 1998).
However, we find that the formation of this feature results from the
adjustment of the initially hydrostatic state to a new dynamic
equilibrium and disappears on the long time scale.

The disk wind accelerates rapidly from the low injection speed to 
fractions of the Keplerian speed.
The flow starts already super Alfv\'enic due to the weak dipolar field
in the disk.
Thus, magneto-centrifugal acceleration along the inclined 
dipole type field lines of the initial magnetic field is
{\em not} the acceleration mechanism.
The acceleration mechanism is mainly due to the centrifugal 
force on the disk matter reaching the non-rotating corona.
Higher above the disk also the Lorentz force contributes to the
acceleration.

The rotating stellar magnetosphere generates a stellar wind.
Due to the strong stellar field, the flow starts
sub-Alfv\'enic.
It is initially magneto-centrifugally driven with a roughly 
spherical Alfv\'en surface.

\subsection{The long-term evolution}
The total mass flow rate into the corona determines how fast 
the flow will establish a (quasi-)stationary state.
First, the outflow evolution is highly time-variable and turbulent. 
After relaxation of the MHD configuration from the initial
magnetohydrostatic state into a new {\em dynamical} equilibrium, 
we finally observe a two-component outflow from disk and star 
distributed smoothly over the whole hemisphere and moving in
spherically radial direction.
After about 2000 rotations a {\em quasi-stationary}
outflow is established over the whole grid (Fig.\,1).
However, small scale instabilities can be observed 
along the neutral field line separating stellar and disk wind 
and along the symmetry axis.

Figure 2 shows the poloidal velocity vectors.
High velocities ($>v_{\rm K,i}$) are only observed far from
the axis.
The asymptotic speed is about $1.5\,v_{\rm K,i}$ for both
components.
The initial axial jet feature disappears.
The axial flow moves with $0.2\,v_{\rm K,i}$.

The axial blobs (or rather tori) generated in this simulation run
move with pattern speed of about 0.1 the Keplerian speed at $r_i$.
We emphasize that the knot size and time scale of knot formation 
in our simulation
is far from the jet knots observed in protostellar jets

The quasi-stationary two-component outflow obtained in our
simulations show almost {\em no indication for collimation}.
This is in agreement with the analysis of Heyvaerts \& Norman (1989)
who have shown that only jets carrying a net poloidal current will
collimate to a cylindrical shape.
In our case we have an initially dipolar magnetosphere
and the final state of a spherically radial outflow enclosing a 
neutral line with a poloidal magnetic field reversal. 
The toroidal field reversal implies a reversal also of the electric 
current and, thus, only a weak {\em net} poloidal current. 
Thus, the flow self-collimation naturally obtained in a
{\em monotonous} magnetic flux distribution
(Ouyed \& Pudritz 1997, Fendt \& Cemeljic 2002) cannot be
achieved.

\end{article}
\end{document}